\documentclass[sigconf]{acmart}

\usepackage{hyperref}
\usepackage[htt]{hyphenat} 

\usepackage{tabularx}
\usepackage{amsfonts} 
\usepackage{multirow}
\usepackage[usestackEOL]{stackengine} 
\usepackage{cellspace}
\usepackage{xcolor}
\usepackage{colortbl}
\newcommand{\grayline}{\arrayrulecolor{lightgray}\hline\arrayrulecolor{black}}

\usepackage{pifont}

\usepackage{makecell}

\usepackage{graphicx}
\graphicspath{ {./images/} }
\usepackage{pifont}
\usepackage{balance}
\usepackage{enumitem}
\usepackage{makecell}
\usepackage{mdframed}

\usepackage{mathtools}
\usepackage{booktabs}
\usepackage{amsmath}
\usepackage{bbm}

\newcommand{\miniskip}{\vspace*{-.9\baselineskip}}
\newcommand{\shrink}{\vspace*{-.9\baselineskip}}

\AtBeginDocument{%
  \providecommand\BibTeX{{%
    \normalfont B\kern-0.5em{\scshape i\kern-0.25em b}\kern-0.8em\TeX}}}

\copyrightyear{2024}
\acmYear{2024}
\setcopyright{rightsretained}
\acmConference[SIGIR '24]{Proceedings of the 47th International ACM SIGIR Conference on Research and Development in Information Retrieval}{July 14--18, 2024}{Washington, DC, USA}
\acmBooktitle{Proceedings of the 47th International ACM SIGIR Conference on Research and Development in Information Retrieval (SIGIR '24), July 14--18, 2024, Washington, DC, USA}
\acmDOI{10.1145/3626772.3657815}
\acmISBN{979-8-4007-0431-4/24/07}

\usepackage{etoolbox}
\makeatletter
\patchcmd{\maketitle}{\@copyrightpermission}{
  \begin{minipage}{0.3\columnwidth}
    \href{http://creativecommons.org/licenses/by/4.0/}{\includegraphics[width=0.90\textwidth]{cc_by4acm.png}}
  \end{minipage}\hfill
  \begin{minipage}{0.7\columnwidth}
    \href{http://creativecommons.org/licenses/by/4.0/}{This work is licensed under a Creative Commons Attribution International 4.0 License.}
  \end{minipage}

  \vspace{5pt}
}{}{}

\makeatother

\begin{document}

\title{Doing Personal LAPS: LLM-Augmented Dialogue Construction for Personalized Multi-Session Conversational Search}


\author{Hideaki Joko}
\affiliation{%
  \institution{Radboud University}
  \country{}
}
\email{hideaki.joko@ru.nl}

\author{Shubham Chatterjee}
\affiliation{%
  \institution{University of Edinburgh}
  \country{}
}
\email{shubham.chatterjee@ed.ac.uk}

\author{Andrew Ramsay}
\affiliation{%
  \institution{University of Glasgow}
  \country{}
}
\email{andrew.ramsay@glasgow.ac.uk}

\author{Arjen P. de Vries}
\affiliation{%
  \institution{Radboud University}
  \country{}
}
\email{a.devries@cs.ru.nl}

\author{Jeff Dalton}
\affiliation{%
  \institution{University of Edinburgh}
  \country{}
}
\email{jeff.dalton@ed.ac.uk}

\author{Faegheh Hasibi}
\affiliation{%
  \institution{Radboud University}
  \country{}
}
\email{f.hasibi@cs.ru.nl}


\renewcommand{\shortauthors}{Hideaki Joko et al.}


\begin{abstract}
The future of conversational agents will provide users with personalized information responses.
However, a significant challenge in developing models is the lack of large-scale dialogue datasets that span multiple sessions and reflect real-world user preferences. Previous approaches rely on experts in a wizard-of-oz setup that is difficult to scale, particularly for personalized tasks.
Our method, LAPS, addresses this by using large language models (LLMs) to \textit{guide} a single human worker in generating personalized dialogues. This method has proven to speed up the creation process and improve quality.
LAPS can collect \textit{large-scale, human-written, multi-session, and multi-domain} conversations, including extracting user preferences.
When compared to existing datasets, LAPS-produced conversations are as natural and diverse as expert-created ones, which stays in contrast with fully synthetic methods.
The collected dataset is suited to train preference extraction and personalized response generation.
Our results show that responses generated explicitly using extracted preferences better match user's actual preferences, highlighting the value of using extracted preferences over simple dialogue history.
Overall, LAPS introduces a new method to leverage LLMs to create realistic personalized conversational data more efficiently and effectively than previous methods.

\end{abstract}

%
%

\ccsdesc[500]{Information systems~Users and interactive retrieval}

%
\keywords{Personalization, Conversational Search, Dialogue Collection}

\maketitle

\section{Introduction}
\label{sec:introduction}

\begin{figure}[t]
\centering
\includegraphics[width=0.99\linewidth]{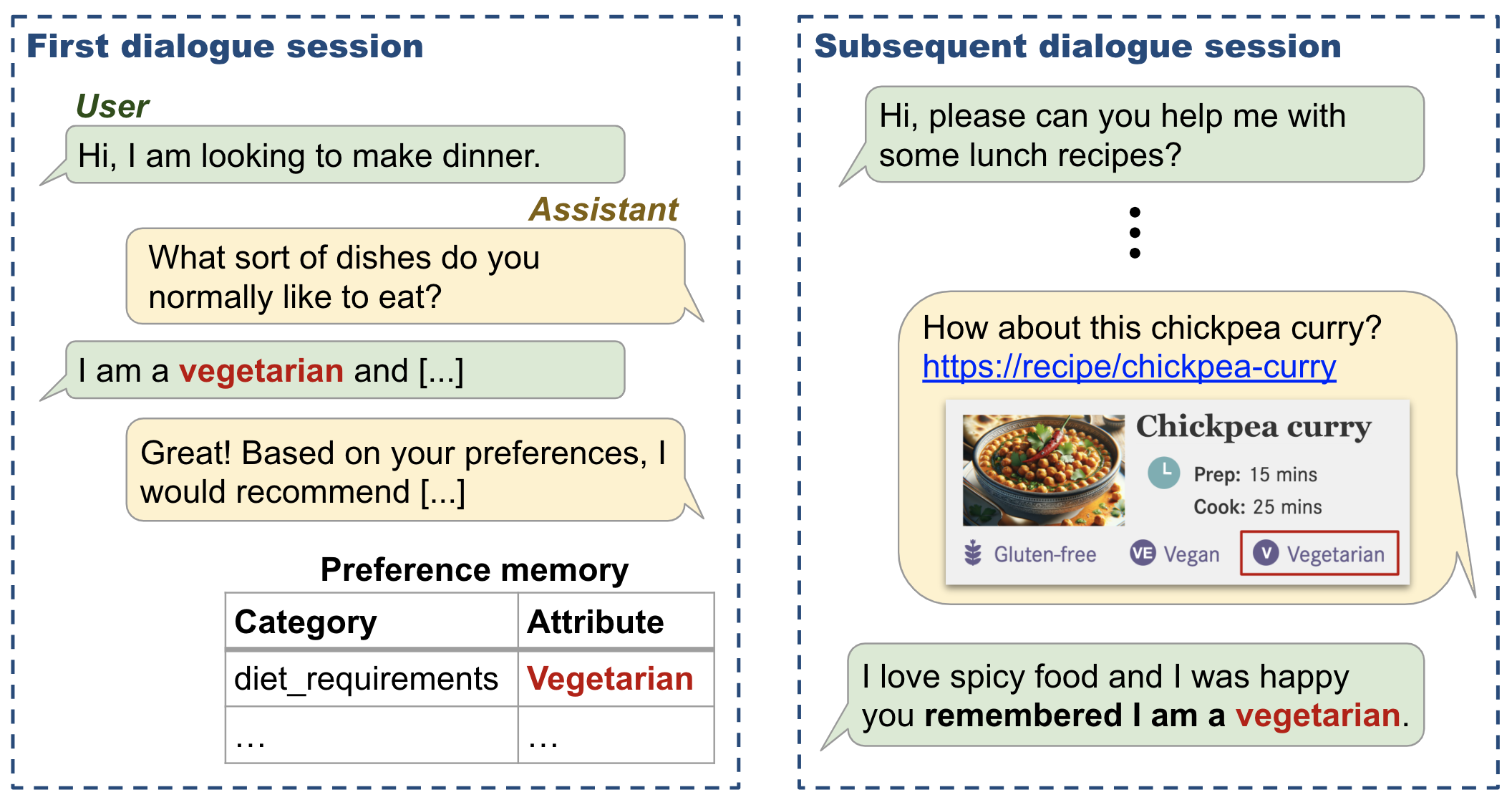}
\shrink
\caption{A snippet from a multi-session dialogue. 
User preferences are extracted and stored in memory to generate personalized recommendations in subsequent sessions.
}
\label{fig:dialogue-illustration}
\miniskip
\end{figure}

Personalization is paramount for conversational search and recommendation~\cite{Luo:2019:LPE, Brown:1987:PSU}. 
Conversational agents need to meet users' expectations and provide them with individually tailored responses. 
In real-world scenarios, where users interact with dialogue agents across multiple sessions, conversational systems need to accurately understand, extract, and store user preferences and articulate personalized recommendations based on the stored user profile; see Figure~\ref{fig:dialogue-illustration}. 
The Information Retrieval (IR) community has studied various aspects of personalization in conversational systems. For instance, the concept of Personal Knowledge Graph (PKG)~\cite{Balog:2019:PKG} is introduced to enable personalization of (conversational) search systems. Similarly, TREC Interactive Knowledge Assistance Track (iKAT) utilizes Personal Text Knowledge Base (PTKB)~\cite{aliannejadi2024trec} for persona-based conversational search.
Recent years have also witnessed tremendous progress in Large Language Models (LLMs)~\cite{Brown:2020:LMF,Ouyang:2022:TLM}. Yet LLM-based conversational agents do not effectively handle user preferences, delivering generalized recommendations that fail to capture the nuanced interests of individual users. 

A major hindrance in developing a personal conversational system is the unavailability of \emph{large-scale human-written} conversational datasets.
These are needed for training new models and understanding user behavior, in expressing their preferences over multiple dialogue sessions~\cite{Radlinski:2019:CCP, Joho:2017:FIW}. However, constructing such datasets has proven a daunting task~\cite{Bernard:2023:MGS, Radlinski:2019:CCP}. 
Preference elicitation in conversations is complex, and crowd workers engage poorly with the task.
While human experts deliver quality results, recruiting them as intermediary coaches or as human agents to generate conversations does not scale.
The recently-emerged paradigm of collecting synthetic conversational data through LLMs~\cite{Gilardi:2023:COC,Yunxiang:2023:CMC,Lee:2022:GPD,Xu:2023:BOC,Chen:2023:PLM,Kim:2023:SMD,Kim:2022:BTF,Jandaghi:2023:FPC} raises concerns regarding the diversity of the generated dialogues~\cite{Reif:2023:VLD,Chung:2023:IDM,Yu:2023:LLM,DellAcqua:2023:NAJ,Park:2023:DDS}.
Crucially, LLM-generated conversations do not represent actual user preferences and interactions, which undermines their credibility for the development of \emph{future} personal conversational systems.

The critical question that arises here is \textit{\textbf{RQ1:} Can we collect large-scale multi-session human-written conversational datasets that contain user preferences?}
We address this question by proposing LAPS, an \underline{\textbf{L}}LM-\underline{\textbf{A}}ugmented \underline{\textbf{P}}ersonalized \underline{\textbf{S}}elf-Dialogue method to collect large-scale personal conversations.
The method employs an LLM to dynamically generate personal \textit{guidance} for crowd workers, playing both user and assistant roles. 
The guidance is generated based on the previously elicited user preferences and the current state of the dialogue, determined by a dialogue act classifier. After each dialogue session, LAPS extracts preferences from the dialogue and stores them in a \textit{preference memory}. This memory is a key-value store about user preferences, analogous to the PKG~\cite{Balog:2019:PKG} and PTKB~\cite{aliannejadi2024trec} concepts.  We show that using LAPS, we can collect 1,406 multi-domain multi-session dialogues, paired with 11,215 preferences.

Our next research question concerns the quality of LAPS-gen\-er\-ated datasets: \textit{\textbf{RQ2:} How do the LLM-augmented self-dialogues compare to human- and synthetically-gen\-er\-ated conversations?} 
We compare our collected conversations with a wide range of widely used conversational datasets and show that LAPS-generated conversations score higher with respect to diversity (based on Dist-n, Ent-n, and SELF-BLEU metrics) and overall quality (based on UniEval~\cite{Zhong:2022:TUM}). We further compare LAPS- and LLM-generated dialogues using GPT-3.5 and GPT-4 and show that LLM-generated dialogues are noticeably less diverse than those involving human workers, even with temperature tuning. 

Although LAPS extracts preferences in a semi-structured format and stores them in a preference memory, one could wonder whether such memory is needed, given LLMs' abilities in handling long context from the previous sessions.
This leads us to the third research question: \textit{\textbf{RQ3:} How can preference memory enhance the effective utilization of user preferences in recommendations?}
To address this question, 
we train a preference extraction model on our dataset and use it to build a preference memory from previous sessions.
These preferences are then incorporated into the LLM's prompt for generating personalized recommendations. 
We compare this approach to the baseline prompting method, where dialogue histories of all sessions are appended to the prompt.
Our experiments show that by incorporating preference memory, the model can more accurately utilize the users' disclosed preferences for recommendations than the baseline method.
The notable advantage of preference memory is that it contributes to more explainable recommendations.
Finally, we found that when using the baseline method, the LLM struggles with recalling user preferences; likely due to lengthy prompts. 

\medskip 
\noindent\textbf{Contributions} of this work are as follows:%
\begin{itemize}[leftmargin=2em]
  \item We introduce LAPS method for collecting scalable multi-session personalized dialogues with actual user preferences.
   
  \item We analyze and compare various dialogue collection methods, demonstrating that LAPS collects lexically diverse and high-quality dialogues, uncovering the diversity issue of generating fully-synthetic dialogues with LLMs.
  \item We study the benefits of storing and using user preferences in a semi-structured format (preference memory) and show that it helps an LLM in recalling previously disclosed preferences when generating personalized recommendations.
   \item Enabled by LAPS, we release a unique conversational dataset that is multi-session, human-written, large-scale, and contains users' personal preferences.\footnote{The code and dataset is available at \url{https://github.com/informagi/laps}}

\end{itemize}

\begin{figure}[t]
  \centering
  \includegraphics[width=\columnwidth]{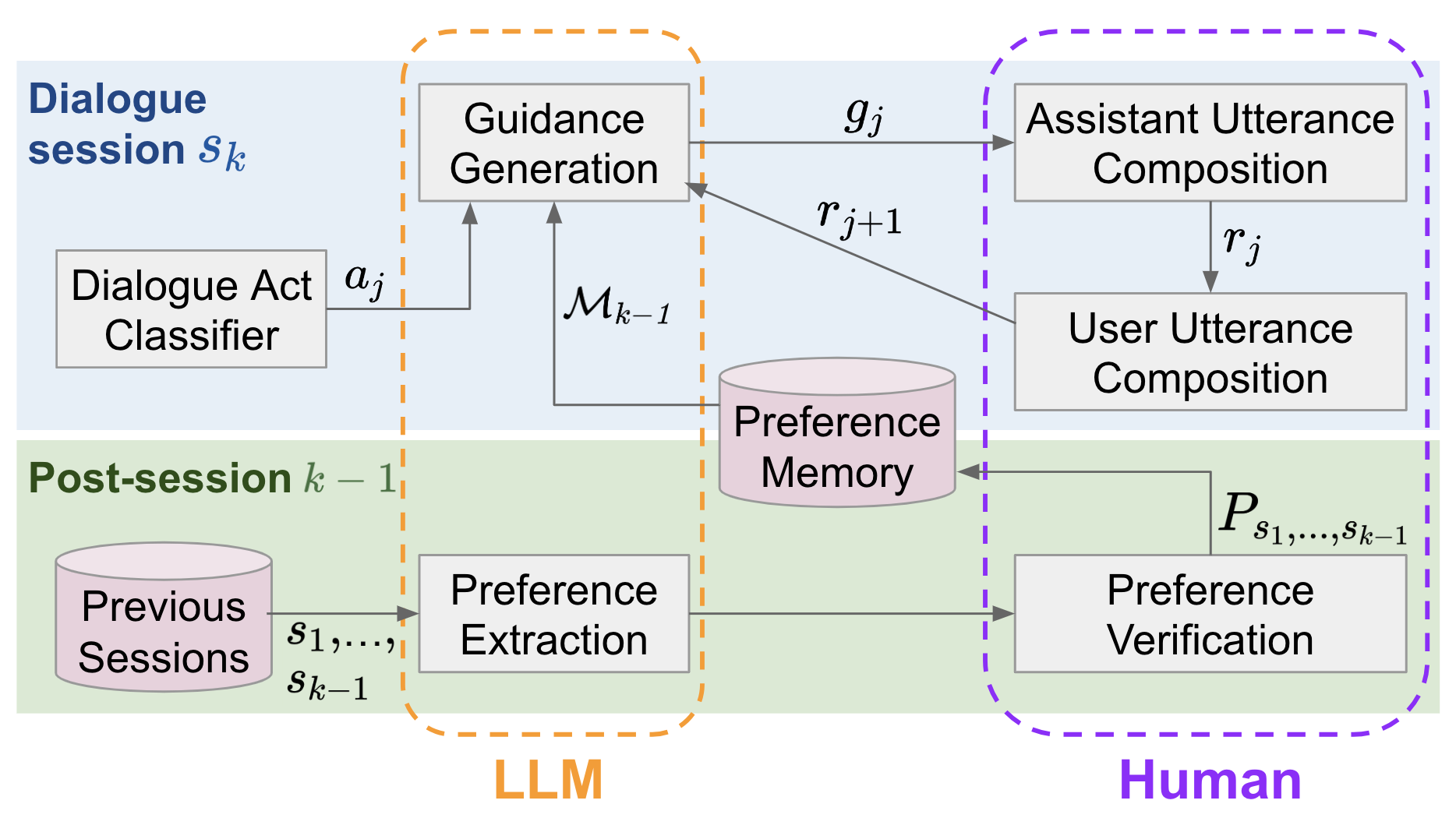}
  \caption{Overview of our dialogue collection method (LAPS).} 
  \label{fig:block-diagram}
\end{figure}
\section{Related Work}
\label{sec:related}

\subsection{Dialogue Collection Methods}
\label{sec:related:collection}

\textbf{Human-Human} interactions are arguably the optimal strategy for collecting natural dialogues. MultiWOZ \cite{Budzianowski:2018:MWO} and PersonaChat \cite{Zhang:2018:PDA} are notable examples, offering task-oriented and chit-chat datasets, respectively. These datasets, however, focus less on real user preferences, relying instead on predefined tasks or personas.
Addressing this limitation, \citet{Radlinski:2019:CCP} introduced CCPE-M, a dataset emphasizing actual user preferences. Similarly, \citet{Bernard:2023:MGS} introduced MG-ShopDial, an e-commerce conversational dataset with genuine preferences. Despite their quality, the small size of these datasets (502 dialogues for CCPE-M and 64 for MG-ShopDial) limits their utility for training large models, and they overlook the multi-session aspect of real-world dialogues.

A quality-quantity trade-off in dialogue collection is highlighted in \cite{Bernard:2023:MGS}. 
Initially attempting crowdsourcing, \citet{Bernard:2023:MGS} shifted to a volunteer-based collection due to low engagement, resulting in higher quality but fewer dialogues. This underscores the challenge in gathering large, high-quality datasets representing true user preferences.
\medskip \noindent \textbf{Self-Dialogue}, where a single worker simulates both roles, is an effective approach for large-scale data collection. Introduced by \citet{Krause:2017:EBO}, this technique has been shown to produce high-quality data, with \citet{Fainberg:2018:TMS} noting its increased coherence and reduced errors compared to human-human dialogue. \citet{Byrne:2019:TTR} further validated this, emphasizing the superior linguistic diversity and fewer mistakes in self-dialogue data.
However, self-dialogue has limitations, especially in managing complex conversations, such as those involving preference elicitation, while acting in two distinct roles.
Additionally, self-dialogue cannot authentically capture unknown user preferences as the same person plays both user and assistant roles, leading to fewer clarification scenarios than in human-human interactions~\cite{Fainberg:2018:TMS}.
Our approach introduces LLM-augmented personalized self-dialogues, leveraging LLMs to ease the cognitive load on workers and addressing the limitation of capturing unknown preferences by involving an LLM, which doesn't have pre-existing knowledge of user preferences.

\medskip \noindent \textbf{(Semi)-Synthetic Dialogue Generation} offers an alternative to relying on crowd workers \cite{Soudani:DAC:2023, Lee:2022:GPD,Xu:2023:BOC,Chen:2023:PLM,Kim:2023:SMD,Kim:2022:BTF,Jandaghi:2023:FPC,Leszczynski:2023:TWS}. \citet{Lee:2022:GPD} developed PERSONACHATGEN using two LLMs for dialogues between personas, requiring multiple model calls for one dialogue. \citet{Chen:2023:PLM} optimized this by using a single model call with a carefully crafted prompt. \citet{Leszczynski:2023:TWS} took a different approach, generating synthetic dialogues by combining user-generated content and metadata.
However, concerns exist about the diversity in LLM-generated texts. \citet{Reif:2023:VLD} noted lexical and syntactic repetition, developing LinguisticLens for syntactic diversity analysis. \citet{Chung:2023:IDM} emphasized the need for human intervention to enhance diversity, and \citet{Yu:2023:LLM} pointed out the uniformity in LLM outputs from simple prompts, suggesting diverse prompts for more varied data. The diversity issue is also evident in fields like social science and business \cite{DellAcqua:2023:NAJ,Park:2023:DDS}. 

To address this, semi-synthetic data collection methods like having crowd workers edit LLM-generated texts have been effective \cite{Shah:2018:BNC,Rastogi:2020:TSM,Bae:2022:BRS}. \citet{Shah:2018:BNC} introduced M2M, a framework using templates for initial dialogue generation and crowd workers for rewriting. Similarly, \citet{Rastogi:2020:TSM} used a schema-guided method. 
Our research improves upon these by using LLMs for providing workers with guidance for response composition, promoting diversity and reducing the influence of generated drafts.

\subsection{Personalized Conversational Systems}
Problem-driven conversational systems, especially those for search and recommendation, benefit from personalization.
Shifting from traditional approaches such as collaborative filtering, recent personalization focuses on more interactive approaches, such as preference elicitation~\cite{Radlinski:2019:CCP,Li:2023:EHP,Christakopoulou:2016:TCR,Sun:2018:CRS,Zhao:2022:KCP,Kostric:2021:SUP,Kostric:2023:GUQ}.
These elicited preferences can be stored as a knowledge graph~\cite{Balog:2019:PKG} through entity linking~\cite{Joko:2021:CEL,Joko:2022:PEC, vanHulst:2020:REL} or in text format~\cite{aliannejadi2024trec}, and later used for personalization. 
Storing user preferences in a (semi-)structured format enables conversational agents to better satisfy users' information needs and can be also useful to mitigate bias\cite{Gerritse:2020:BCS}. 
Following this line, our method elicits user preferences and stores them in a semi-structured preference memory.

\medskip \noindent \textbf{Memory and feedback} also offer the promise of making systems better aligned with user needs.
This approach, tracing back to ALFRED~\cite{Riesbeck:1981:FDR}, involves storing past failures to improve future interactions. Recently, \citet{Madaan:2023:SIR} advanced this concept with SELF-REFINE, enabling LLMs to refine their outputs iteratively using their own feedback. 
For tasks requiring deeper personalization, human feedback is invaluable.
\citet{Madaan:2022:MPE} enhanced LLM response quality by adding a memory module that remembers information from the user's past sessions.
Aligned with \cite{Madaan:2022:MPE}, our approach includes a memory module to store user preferences from past interactions, enabling enhanced personalization.
\section{Dialogue Collection Method}
\label{sec:method}

{\renewcommand{\arraystretch}{1.3}
\begin{table*}[t]
\shrink
\centering
\footnotesize
\setlength{\tabcolsep}{3pt} 

\caption{Overview of the selected baseline datasets and ours. EXP and CW denote an expert and a crowd worker, respectively.} 

\label{tab:dataset-comparison}
\begin{tabular}{l || l | c | l | l | c | c | c | c | c }
\hline
\multirow{2}{*}{\textbf{Dataset}} & \multicolumn{1}{l|}{\textbf{Collection}} & \multirow{2}{*}{\textbf{\#Dial}} & \multirow{2}{*}{\textbf{Tasks}} & \multirow{2}{*}{\textbf{Domains}} & \multirow{2}{*}{\textbf{Scalability}} & \multicolumn{1}{c|}{\textbf{Actual User}} & \multicolumn{1}{c|}{\textbf{Preference}} & \multicolumn{1}{c|}{\textbf{Preference}} & \multicolumn{1}{c}{\textbf{Multi-}} \\ 
& \multicolumn{1}{l|}{\textbf{Method}} & & & & \multicolumn{1}{c|}{\textbf{}} & \multicolumn{1}{c|}{\textbf{Preferences}} & \multicolumn{1}{c|}{\textbf{Elicitation}} & \multicolumn{1}{c|}{\textbf{Extraction}} & \multicolumn{1}{c}{\textbf{Session}} \\
\Xhline{2pt}
\makecell[l]{SGD~\cite{Rastogi:2020:TSM}} & \makecell[l]{Semi-synthetic} & 16,142 & \makecell[l]{Booking, rec., etc.} & \makecell[l]{20 domains, inc. restaurants} & \ding{51} & \texttimes & \texttimes & \texttimes & \texttimes \\
\grayline
\makecell[l]{M2M~\cite{Shah:2018:BNC}}  & \makecell[l]{Semi-synthetic} & 3,008 & \makecell[l]{Booking} & \makecell[l]{Restaurants, movies} & \ding{51} & \texttimes & \texttimes & \texttimes & \texttimes \\
\grayline
\makecell[l]{PersonaChatGen~\cite{Lee:2022:GPD}} & \makecell[l]{Fully-synthetic} & 1,649 & \makecell[l]{Personal chit-chat} & \makecell[l]{Open domain} & \ding{51} & \texttimes & \texttimes & \texttimes & \texttimes \\
\grayline
\makecell[l]{Taskmaster-1~\cite{Byrne:2019:TTR}} & \makecell[l]{Self-dialogue} & 7,708\textsuperscript{\dag} & \makecell[l]{Booking, ordering} & \makecell[l]{6 domains, inc. restaurants}  & \ding{51} & \texttimes & \texttimes & \texttimes & \texttimes \\
\grayline
\makecell[l]{MultiWOZ~\cite{Budzianowski:2018:MWO}}  & \makecell[l]{Human-human (CW)} & 8,438 & \makecell[l]{Booking, rec., etc.} & \makecell[l]{7 domains, inc. restaurants} & \ding{51} & \texttimes & \texttimes & \texttimes & \texttimes \\
\grayline
\makecell[l]{CCPE-M~\cite{Radlinski:2019:CCP}}  & \makecell[l]{Human-human (EXP)} & 502 & \makecell[l]{Rec.} & \makecell[l]{Movies} & \texttimes & \ding{51} & \ding{51} & $\triangle$ & \texttimes \\
\grayline
\makecell[l]{MG-ShopDial~\cite{Bernard:2023:MGS}} & \makecell[l]{Human-human (EXP)} & 64 & \makecell[l]{Rec., QA, etc.} & \makecell[l]{E-commerce} & \texttimes & \ding{51} & \ding{51} & \texttimes & \texttimes \\
\hline
\makecell[l]{LAPS} & \makecell[l]{LAPS} & 1,406 & \makecell[l]{Rec.} & \makecell[l]{Recipes, movies} & \ding{51} & \ding{51} & \ding{51} & \ding{51} & \ding{51} \\
\hline
\end{tabular}
\tiny
\begin{flushleft}
\hspace{1.0em} \textsuperscript{\dag} Includes only self-dialogues.\\
\hspace{1.0em} $\triangle$ Annotations are provided but no entity-relation pair extraction with like/dislike distinction.
\end{flushleft}
\end{table*}}

We propose LAPS, an LLM-Augmented Personalized Self-Dialogue construction method, capable of collecting large-scale, human-written, multi-session, and multi-domain conversations, paired with extracted user preferences. The method consists of four key elements (cf. Fig.~\ref{fig:block-diagram}): (i) dialogue act classification, (ii) guidance generation, (iii) utterance composition, and (iv) preference extraction.

The dialogue act classifier determines the next action that the assistant should take; e.g., recommend.  
Based on the dialogue act, the LLM generates guidance considering the dialogue history and the previously extracted preferences.
The human agent then composes the assistant response based on the LLM-generated guidance, and then switches to the role of a user, providing a response to the previous utterance. The process continues until a relevant recommendation is made and the dialogue session is completed.
Upon completion of a session, the preferences are extracted from the dialogue using an LLM and checked by the human agent.
These preferences, once confirmed by the same human agent, are stored in the \textit{preference memory} and used in subsequent sessions for generating personalized guidance. The human agent is then encouraged to initiate a new dialogue session for another scenario in the given domain.  This process continues until the human agent exits the job or reaches the end of all pre-defined session scenarios. 

%

\subsection{Task Formulation}
\label{sec:method:formulation}
The objective of this task is twofold: firstly, to create a large-scale collection of multi-session dialogues written by humans, focusing on user preferences; and secondly, to extract and compile the specific user preferences mentioned within these dialogue sessions.
Formally, the dialogue collection method $F$ is defined as a mapping from a set of task descriptions $\mathbb{T}$ and human agents $\mathbb{H}$ to a set of dialogue sessions and their extracted preferences $\mathbb{S}$:
\begin{align*}
    &F : \mathbb{T}, \mathbb{H} \rightarrow \mathbb{S}\\
    &F(t,h) = [(s_1, P_{s_1}), \ldots, (s_n, P_{s_n})],
\end{align*}
%
where $t$ is a task description for a topic,  $h$ is a human agent with identical preferences for a given topic, and $n$ represents the total number of sessions. The dialogue session $s_i$ and its corresponding preference set $P_{s_i}$ are defined as:
%
%
\begin{align}
    &s_i = [u_1^i, u_2^i, \ldots, u_m^i], \nonumber \\
    &P_{s_i} = \{(c, p_j)   \, | \, c \in C, j \in \mathbb{N}\}, \label{eq:pref}
\end{align}
where the dialogue session $s_i$ composed of a sequence of utterances $u$, and the preference set $P_{s_i}$ is a set of  category-preference pairs $(c, p_j)$, where the category $c$ belongs to the set of categories $C$; e.g., \{\texttt{allergy}, \texttt{cuisine}, \texttt{diet}\}.
We note that the preference set $P_{s_i}$ is generated only after the completion of session $s_i$, by extracting user preferences from the dialogue post-session and validating them with the same worker. The extracted preferences of a human agent are stored in a memory component $\mathcal{M}$, defined as: 
%
\begin{equation*}
\mathcal{M}_k = \bigcup_{\substack{1 < i \le k \\ k \le n}} P_{s_i},
\end{equation*}
where session $s_k$ is the last completed session by the human agent. 

Here we draw an analogy between the preference memory $\mathcal{M}$ and Personal Knowledge Graphs (PKGs) ~\cite{Balog:2019:PKG}, where the personal information of users is stored according to an ontology. The \texttt{<subject, predicate, object>} triplets in PKGs correspond to human ($h$), category ($c$), and preference ($p$) in preference memory, respectively. We note that unlike a PKG that is built based on a pre-defined ontology, preference memory uses a more relaxed version of categories that are extracted on-the-fly from user utterances.  Preference memory can be also viewed  as a semi-structured form of PTKB, where free-form sentences about a user's persona are transformed in a key-value format.
\subsection{Guidance Generation}
Generating guidance for human agents is central to collecting large-scale, high-quality,  human-written utterances in LAPS.
Large-scale construction of conversational data requires recruiting crowd workers. However, due to the complexity of the task and high cognitive load of generating conversational utterances, crowd workers show poor engagement~\cite{Bernard:2023:MGS}. The challenge is even more intense for the preference elicitation task, where we need to coach the human agent simulating the system to ask engaging questions to reveal user preferences~\cite{Radlinski:2019:CCP}. A remedy could be utilizing LLMs to generate system utterances. This, however, results in less diverse conversations and is not in line with our aim of generating training data for \emph{future} personalized conversational search and recommendation systems. Even by instructing crowd workers to re-write LLM-generated utterances, we observed (in our pilot studies) that crowd workers become less creative in generating their own utterances and tend to replicate pre-generated utterances.

To alleviate the aforementioned problems, we propose to coach crowd workers throughout the conversation using automatically generated personalized guidance, and let the workers compose their own utterances via dialogue self-play. 
Using this approach, we reduce the cognitive load of a highly complex task to a minimum, allowing workers to focus on a simple sub-task at a time and generate high-quality and engaging conversations. Formally, to compose the assistance utterances $u_j$, the human agent receives personalized guidance $g_j$ generated by an LLM. The guidance generation function $\mathcal{G}$ is defined as:
\begin{equation*}
\mathcal{G}\left(u_1, \ldots, u_{j-1}, a_j, \mathcal{M}_k\right) = g_{j},
\end{equation*}
%
where $\mathcal{M}_k$ denotes the preference memory extracted from sessions $(s_1, \ldots, s_k)$, utterances $u_1, \ldots, u_{j-1}$ represents conversations history up until turn $j$, and $a_j$ is the action that needs to be taken for turn $j$, obtained from the dialogue act classifier (cf. \S~\ref{sec:method:act}).


\medskip \noindent \textbf{Instantiation.}
As an instantiation of this function, we prompt GPT-3.5 \texttt{turbo} to generate personalized guidance. The guidance prompt template takes the dialogue history $u_1, \ldots, u_{j-1}$, preference memory $\mathcal{M}_k$, and instructions for the current dialogue act $a_j$ as inputs. The prompts include detailed step-by-step instructions for chain-of-thought prompting~\cite{Wei:2022:CTP}.
For instance, the prompt for the \emph{preference elicitation} act in a given \texttt{DOMAIN} is as follows:
\vspace{0.5em}
\begin{mdframed}[linecolor=black, linewidth=0.5pt, roundcorner=5pt, backgroundcolor=gray!10, font=\small]
    \textit{You are an advisor, who supports} \textbf{\$\texttt{DOMAIN}} \emph{recommendation assistants to compose responses to users.
    In this step, the assistant \textbf{collects information about user's preferences}} ([$a_j$]).
    
    \vspace{2mm}
    
    \noindent Preference memory: [$\mathcal{M}_k$]\\
    \noindent Dialogue history: [$u_1, \ldots, u_{j-1}$]
    
    \vspace{2mm}
    \noindent \textit{Step 1: Identify the last user turn.\\
    Step 2: Explain the intent of the last user utterance.\\
    Step 3: \textbf{Which preference(s) should the assistant ask next} given the dialogue history?\\
    Step 4: \textbf{Compose very short guidance for the human assistant on how to write a response to the user.}\\
    Step 5: Output in JSON format following [...]\\
    Ensure that you distinctly label and delineate Steps 1, 2, 3, 4, and 5. Let's think step by step:
    }
\end{mdframed}
\vspace{0.5em}
The guidance prompt for \emph{recommendation} act is similar to the \emph{preference elicitation} act, except that the assistant is instructed to recommend an item based on the user's personal preferences with URLs. The guidance also includes ``\textit{When making recommendations, if necessary, effectively utilize the user's preferences, such as [...]}'' to encourage the assistant to use the user's preferences disclosed in the previous sessions if necessary.

\subsection{Dialogue Act Classification}
\label{sec:method:act}
Dialogue act is an action that can change the (mental) state of conversation and guide the system to generate the next utterance~\cite{Gao:2019:NAC}. Dialogue act is used as an input to the guidance generation function and is obtained by the dialogue act classifier $\mathcal{A}$:
%
\begin{equation*}
    \mathcal{A}(u_{1}, \ldots, u_{j-1}, a_{j-1}) = a_{j}
\end{equation*}
which determines action $a_j$ based on the dialogue history $u_1, \ldots, u_{j-1}$ and the previous dialogue act $a_{j-1}$.


\medskip \noindent \textbf{Instantiation.}
We instantiate the act classifier function by prompting GPT-3.5 \texttt{Turbo}.
A series of dialogue acts are defined, outlining the specific actions to be taken sequentially. The primary dialogue acts in our setup are: (1) \textit{greeting}, (2) \textit{preference elicitation}, (3) \textit{recommendation}, (4) \textit{follow-up questions}, and (5) \textit{goodbye}. A dialogue act is selected upon completion of the previous act, which is determined by the LLM using detailed chain-of-thought instruction prompts; e.g., the \textit{recommendation} act is selected when the LLM produces the response ``true'' for the instruction prompt ``\textit{[...] Has the user shared any of the preferences listed above? Has the assistant collected why the user has the preference? If both are true, return true.}''



When collecting preferences for the first session, the \textit{preference elicitation} act is further divided into three sub-actions for collecting (i) \emph{must-have}, (ii) \emph{should-have}, and (iii) \emph{could-have} preferences. Once the \emph{must-have} and \emph{should-have} preferences are collected, the subsequent sessions only collect \emph{could-have} preferences, using the preference memory for other preferences.  

\subsection{Utterance Composition}
\label{sec:method:utterance-composition}
Conversation utterances are composed by the human agent for both user and assistant roles via dialogue self-play.
For the assistant role, the composition process is supported by the LLM-generated \textit{guidance}, which enables generating diverse human-written system utterances. For the user role, the human agent mainly needs to elaborate his preferences and state his opinion about the recommendations. The system and user utterance generation functions are defined as:
%
\begin{align*}
 & \mathcal{W}_s\left(S_k, g_{j}, u_1, \dots, u_{j-1}\right) =  u_j, \\
 & \mathcal{W}_u\left(S_k, u_1, \dots, u_j\right) = u_{j+1}, 
\end{align*}
where $\mathcal{W}_s$ and  $\mathcal{W}_u$  represent the function for writing system and user responses, respectively.
Here, $g_j$ denotes the guidance for generating utterance $u_j$, and  $S_k = \{s_1, \dots, s_k\}$ represents the session history, comprising all previous dialogue sessions up until, but not including, the current session.

\medskip \noindent \textbf{Instantiation.}
These functions are instantiated by recruiting crowd workers and instructing them to write a self-dialogue for our task: 
\textit{``Your task is to chat with yourself both as a user (seeking a cooking recipe or movie) and an assistant (offering recipe or movie recommendations). You need to discuss preferences and receive suggestions while playing both roles.''}
This instruction is followed by brief descriptions of roles, session settings, and chat interface instructions, as well as general information about the payment and rejection policy.

For the user role, workers are instructed to be themselves, provide their preferences, review the recommendations, and offer feedback.
In contrast, the assistant role entails more tasks. Assistants must elicit preferences to inform their recommendations and provide URLs for the recommended items. Following user feedback, assistants confirm why the user likes or dislikes the recommendation to obtain a reusable preference for the next session.

\subsection{Preference Extraction}
Upon completion of a dialogue session, user preferences are extracted from the dialogue. Due to the inherent ambiguity and complexity of human-written conversations, we collect these preferences from the same human agent that generates the conversation.
Note that here we only collect user preferences that are mentioned in the course of previous conversation sessions and not general user preferences. Formally, given the dialogue session $s_i$, the preference extraction function $\mathcal{E}$ is defined as:
%
\begin{equation}
\label{eq:preference-extraction}
\mathcal{E}(s_i, C) =  P_{s_i},
\end{equation}
where $C$ denotes the set of preference categories, and $P_{s_i}$ is the set of category-preference pairs extracted from the dialogue session.


\medskip \noindent \textbf{Instantiation.}
Extraction of preferences is performed using a semi-automated approach, where the initial set of preferences is extracted by an LLM and then validated by the human agent. We prompt GPT4 with chain-of-thought instructions to generate preference attributes for each preference category $c \in C$, given the dialogue session $s_i$ and preference categories $C$.


    
    

Once an initial set of preferences is extracted, the worker is instructed to confirm the correctness of each extracted preference and verify all disclosed preferences during the conversations are extracted.
The worker is then directed to start a new conversation session or terminate the task.

\subsection{ Evaluation}
\label{sec:method:dialogue-eval}

\medskip \noindent \textbf{Baseline Datasets.}
We evaluate our LAPS method by comparing existing conversational datasets with the conversations generated by LAPS. 
We carefully select the baseline datasets by examining 170 datasets listed by~\citet{Joko:2021:CEL}, and narrowing down our selection by the following criteria: (i) inclusion of preference elicitation, (ii) utilization of preferences, and (iii) focus on task-oriented dialogues. We prioritize datasets that are both published in peer-reviewed venues and publicly available. 
The selected datasets are summarized in Table~\ref{tab:dataset-comparison}.
For a fair comparison, when possible, we select a single domain from each baseline dataset that overlaps with the domains of the dataset created by LAPS, namely recipe and movie. For open-domain datasets, e.g., PersonChatGen~\cite{Lee:2022:GPD}, we select dialogues with food-related personas, categorized under ``Food'' and ``Drink.''

\medskip \noindent \textbf{Baseline Methods.}
We tried three other human-based dialogue collection methods: \emph{Human-Human}, \emph{self-dialogue}, and \emph{LLM-human}. These methods do not scale and cannot generate quality conversations. In the \emph{Human-Human} method, we encountered difficulties in pairing up two workers simultaneously due to the high dropout rates, caused by the complex nature of multi-session preference elicitation. For the \emph{self-dialogue} method, we followed ~\cite{Speggiorin:2022:TPM} and simplified the assistant role by providing users with a predefined response. This, however, led to homogeneous dialogues, even though the workers were instructed to rewrite the response. In the \emph{LLM-human} method, we generated assistant responses using GPT-3.5 \texttt{turbo} and asked workers to rewrite the utterances. However, even after experimenting with multiple temperature settings, the dialogue remained homogeneous (a common issue with LLMs reported also in~\cite{Reif:2023:VLD, Chung:2023:IDM, Yu:2023:LLM, DellAcqua:2023:NAJ, Park:2023:DDS}) and workers often accepted the grammatically correct responses without re-writing them.

While we focus on collecting dialogues with actual user preferences, one might wonder whether an LLM could also play the role of a human agent. To address this question, we prompt the LLM to act both as a user and an agent, using the same input and output as human agents in LAPS (cf.\S~\ref{sec:method:utterance-composition}). We report on the results obtained from GPT-3.5 \texttt{turbo} and GPT-4-1106 \texttt{preview} with a temperature parameter of 1.0. These dialogues are 3 sessions long.

\medskip \noindent \textbf{Dialogue Diversity.}
We use three metrics to measure the lexical diversity of the collected conversations: (i) Dist-n~\cite{Li:2016:DPO}, (ii) Ent-n~\cite{Zhang:2018:GID}, and (iii) Self-BLEU~\cite{Zhu:2018:TBP}.
\textit{\textbf{Dist-n}} measures the response diversity by computing the ratio of distinct n-grams to all n-grams in the given collection. Following~\cite{Li:2016:DPO}, we use Dist-1 and -2 to measure the lexical diversity of conversation utterances.
\textit{\textbf{Ent-n}} aims to enhance the Dist-n measure by incorporating the frequency differences of n-grams into consideration, leveraging the entropy of the n-gram distribution. We report on Ent-4, following~\cite{Zhang:2018:GID}. 
\textit{\textbf{Self-BLEU}} considers one utterance as a hypothesis and the rest of the utterances in the collection as references and calculates the BLEU score for each utterance. Following the NLTK's default setting~\cite{Bird:2009:NLP} and~\cite{Zhu:2018:TBP}, we compute the BLEU scores for $n=1,2,3,4$ for each hypothesis-reference pair and take the mean over all computed BLEU scores.

\textit{Normalization.}
A frequently overlooked pitfall of diversity metrics is their dependency on the total number of words in the dialogues.
For instance, Dist-n scores are typically higher for datasets with fewer total number of words, as they are less likely to have repeated words.
To address this, we set a word cutoff for all datasets and randomly sample dialogues until the total word count reaches this cutoff.
The cutoff is set at 7,012 words, which is the minimum number of total words for user/system utterances across all datasets. 
We perform 100 random sampling per dataset and report the average scores, as well as two-tailed independent t-test results.
Additionally, since M2M~\cite{Shah:2018:BNC} dataset is lowercased, we lowercase all other datasets for a fair comparison.


\medskip \noindent \textbf{Dialogue Quality.}
Dialogue quality evaluation is inherently challenging due to its subjective nature. 
Human evaluation, often considered the gold standard, is known to be highly sensitive to task design and instructions. Even with much care, it still suffers from differing bias and high variance per annotator, especially in crowdsourced environments~\cite{Deriu:2021:SEM,Smith:2022:HEC,Finch:2023:DFY,Li:2019:AID,Zhang:2023:BLL}.
Automatic evaluation, while capable of mitigating the aforementioned issues, is also known to have its own limitations, including a bias towards machine-generated responses~\cite{Liu:2023:GNE}.
Aware of these limitations, we opt for automatic evaluation for two reasons: (1) our aim is to ensure our dialogue quality aligns with other high-quality datasets, rather than attaining state of the art, and (2) we use human workers to compose responses in their own words, which is less likely to be overestimated by metrics biased towards machine-generated responses.

After examining four reference-free automatic evaluation methods~\cite{Mehri:2020:URF,Lin:2023:UMA,Liu:2023:GNE,Zhong:2022:TUM}, we select \textbf{UniEval}~\cite{Zhong:2022:TUM} as our automatic evaluation metric considering availability, cost, and performance.
For evaluation aspects, we choose naturalness, understandability, and coherence from~\citet{Zhong:2022:TUM}.
The aspects that require the conditioning fact as an input are not used, as the factuality of user preferences falls outside the scope of our study.
We use the official implementation of UniEval\footnote{\url{https://github.com/maszhongming/UniEval}} and its default settings. 
To ensure that the evaluation is computationally feasible using our available computational resources, we randomly select 100 dialogues (consisting of 1.8K responses on average) from each dataset.
For significant testing, we use two-tailed independent t-tests ($p < 0.05$).

\subsection{Experimental Setup}
{\medskip \noindent \textbf{Domains.}
Our domains are \textit{recipe} and \textit{movie}.
The recipe domain involves planning for the next dinner (session 1), breakfast (session 2), and lunch (session 3).
The movie domain involves planning to watch a movie with family, friends, or alone (session 1), exploring another movie by the same director or actress/actor as the previous recommendation (session 2), and watching with different people or a different occasion (session 3). For each domain, we curated a list of categories that are relevant to the task and categorized them into categories of  \textit{must-have}, \textit{should-have}, and  \textit{could-have}. These categories are detailed in our online repository.
}

\medskip \noindent \textbf{Participants and Quality Control}.
We recruited Prolific\footnote{\url{https://www.prolific.com/}} workers from English-speaking countries,
having $ \geq 98\% $ approval rate and $ \geq 1000 $ previous submissions.
For the movie domain, we only invited workers that accurately performed that task for the recipe domain. Throughout the experiments, we actively communicated with workers, answering over 250 questions and incorporating their feedback to clarify instructions.
£14 was paid for completing three sessions, which took \textasciitilde65 minutes.
Workers were also allowed to terminate the task at an earlier session and receive partial payment.
This allowed us to collect high-quality multi-session dialogues, as workers completing all sessions tend to be more engaged in the task.
After crowdsourcing, we manually reviewed the dialogues to ensure data quality, making corrections or deletions as necessary.
Common errors (aside from malicious behavior of not providing meaningful responses) include failing to include URLs in recommendations and misunderstanding their current role in the conversation. 

\medskip \noindent \textbf{Chat Interface.}
We collect conversations using TaskMAD~\cite{Speggiorin:2022:TPM} and further develop it to support new features for our task.
The human agent interacts with two chat interfaces for system and user roles, and each interface consists of a text box for composing responses and an instruction box to clarify role responsibilities.


\begin{table}[t]
\centering
\small
\shrink
\caption{Statistics of LAPS dataset.}
\shrink
\label{tab:dataset-overview}
\setlength{\tabcolsep}{3.5pt}
\begin{tabular}{l | l || c | c | c | c | c | c }
\hline
\multirow{3}{*}{\textbf{Domain}} & \multirow{3}{*}{\textbf{Split}} & \multicolumn{3}{c|}{\textbf{\#Dialogue Sets}} & \multirow{3}{*}{\textbf{\#Pref}} & \multirow{3}{*}{\textbf{\#Utt}} & \multirow{3}{*}{\textbf{\#Dial}} \\ 
\cline{3-5}
& & \textbf{\footnotesize Single-} & \textbf{\footnotesize Two-} & \textbf{\footnotesize Three-} & & & \\[-2.5pt]
& & \textbf{\footnotesize Session} & \textbf{\footnotesize Session} & \textbf{\footnotesize Session} & & & \\[-1pt]
\Xhline{2pt}
\multirow{4}{*}{Recipe} & Train & 163 & 24 & 160 & 5,538 & 9,342 & 691 \\
& Val & 24 & 5 & 21 & 772 & 1,333 & 97 \\
& Test & 48 & 10 & 41 & 1,600 & 2,610 & 191 \\
& \textbf{Total} & \textbf{235} & \textbf{39} & \textbf{222} & \textbf{7,910} & \textbf{13,285} & \textbf{979} \\
\hline
\hline
\multirow{4}{*}{Movie} & Train & 46 & 14 & 72 & 2,225 & 3,974 & 290 \\
& Val & 5 & 1 & 13 & 351 & 642 & 46 \\
& Test & 11 & 4 & 24 & 729 & 1,220 & 91 \\
& \textbf{Total} & \textbf{62} & \textbf{19} & \textbf{109} & \textbf{3,305} & \textbf{5,836} & \textbf{427} \\
\hline
\end{tabular}
\end{table}

\section{Personal Recommendation Method}
\label{sec:recommendation}

\begin{table}[t]
\centering
\caption{Lexical diversity scores.
    Significance against all baselines is marked by \textsuperscript{+}.
}
\shrink
\label{tab:eval-baselines}
\setlength{\tabcolsep}{7pt}
\begin{tabular}{l || ccc }
\hline
\multirow{2}{*}{\textbf{Dataset}} & \multirow{2}{*}{\textbf{Dist-1/2}} & \multirow{2}{*}{\textbf{Ent-4}} & \multicolumn{1}{c}{\textbf{Self-}} \\
& & & \multicolumn{1}{c}{\textbf{BLEU\textsuperscript{$\blacktriangledown$}}} \\
\Xhline{2pt}
SGD & 0.179 / 0.538 & 8.311 & 0.964 \\
M2M & 0.057 / 0.290 & 7.922 & 0.955 \\
PersonaChatGen & 0.165 / 0.523 & 8.261 & 0.970 \\
Taskmaster-1 & 0.207 / 0.644 & 8.384 & \underline{0.949} \\
MultiWOZ & 0.158 / 0.505 & 8.345 & 0.966 \\
CCPE-M & 0.175 / 0.571 & 8.414 & 0.961 \\
MG-ShopDial & \textbf{0.234} / \underline{0.653} & 8.199 & \textbf{0.935} \\
\hline
LAPS-Recipe & 0.207 / 0.650 & \underline{8.563}\textsuperscript{+} & 0.955 \\
LAPS-Movie & \underline{0.222}\textsuperscript{+} / \textbf{0.666}\textsuperscript{+} & \textbf{8.593}\textsuperscript{+} & 0.954 \\
\hline
\end{tabular}
\footnotesize
\begin{flushleft}
\hspace{1.5em} \textsuperscript{$\blacktriangledown$} Lower is better.
\end{flushleft}
\shrink
\end{table}

The end goal for large-scale preference elicitation conversational datasets is to enable personal conversational search and recommendation. Based on LAPS, we collect a large-scale dataset for the movie and recipe domains and use it to train a model for extracting personal preferences from conversations. The preferences are then used to generate personalized recommendations. 




\subsection{Preference Extraction}
\label{sec:rec:extraction}
In this task, we aim to automatically extract user preferences from a dialogue session (cf.\ Eq.~\ref{eq:preference-extraction}).
We cast this task as a seq-to-seq QA~\cite{Chung:2022:SIL} and fine-tune an LLM with instruction prompts eliciting user preference regarding category and conversation session. Formally, our method decomposes Eq.~\ref{eq:preference-extraction} as $ \mathcal{E}(s, \mathcal{C}) = \bigcup_{c \in \mathcal{C}} \mathcal{E}'(s, c) $, where $\mathcal{E}'$ is the preference extraction function for individual category $ c $.



We use FlanT5~\cite{Chung:2022:SIL} as the base model and use instruct prompts to read the given session and answer questions about the user's preferences, such as \textit{``What cuisine does the user like?''}
For the recipe domain, we use FlanT5 Small (80M parameters), Base (250M), and Large (780M) models and fine-tune them on our recipe dataset. For the movie domain, we explore a domain adaptation approach, where the initially fine-tuned model on the recipe domain (FlanT5-Large) is further fine-tuned on the movie domain.



\subsection{Personalized Recommendation}
With the personalized recommendation task, we aim to generate a recommendation response based on user's personal preferences.
Personal preferences are stated in the conversation history $h$ and all previously completed sessions $S_k$. By utilizing the preference extraction method (cf. \S~\ref{sec:rec:extraction}), we construct the preference memory $\mathcal{M}_k$ based on session history $S_k$ and use it for recommendation. Formally, recommendation utterance $ u^{pred} $ is generated by the recommendation generation function $R$, defined as $R\left(h, \mathcal{M}_{k}\right) = u^{pred} $.
Recommendation responses are generated by an LLM (LLaMA-2-7B~\cite{Touvron:2023:LOF}) using zero-shot prompting. The prompt contains instructions to generate a personalized recommendation appended with the preference memory and history of the current conversation. 

\begin{table}[t]
\centering
\caption{Lexical diversity scores of synthetic dialogue generation.
Significance against all baselines is marked by \textsuperscript{+}.
}
\shrink
\setlength{\tabcolsep}{3pt}
\label{tab:eval-synthetic}
\begin{tabular}{l|l||ccc}
\hline
\multirow{2}{*}{\textbf{Domain}} & \multirow{2}{*}{\textbf{Method}} & \multirow{2}{*}{\textbf{Dist-1/2}} & \multirow{2}{*}{\textbf{Ent-4}} & \multicolumn{1}{c}{\textbf{Self-}} \\
& & & & \multicolumn{1}{c}{\textbf{BLEU\textsuperscript{$\blacktriangledown$}}} \\
\Xhline{2pt}
\multirow{3}{*}{Recipe} & Synthetic GPT-3.5 & 0.127 / 0.430 & 8.308 & 0.981 \\
 & Synthetic GPT-4 & 0.183 / 0.597 & \textbf{8.601} & 0.976 \\
 & LAPS & \textbf{0.207}\textsuperscript{+} / \textbf{0.65}\textsuperscript{+} & 8.563 & \textbf{0.955}\textsuperscript{+} \\
\hline
\hline
\multirow{3}{*}{Movie} & Synthetic GPT-3.5 & 0.138 / 0.444 & 8.331 & 0.977 \\
 & Synthetic GPT-4 & 0.178 / 0.559 & 8.481 & 0.979 \\
 & LAPS & \textbf{0.222}\textsuperscript{+} / \textbf{0.666}\textsuperscript{+} & \textbf{8.593}\textsuperscript{+} & \textbf{0.954}\textsuperscript{+} \\
\hline
\end{tabular}
\footnotesize
\begin{flushleft}
\hspace{0.7em} \textsuperscript{$\blacktriangledown$} Lower is better.
\end{flushleft}
\end{table}

\subsection{Evaluation}
\label{sec:downstream:eval-metric}

\medskip \noindent \textbf{Recommendation Baseline.}
 For our personalized recommendation method, we consider a baseline method, where the preference memory $\mathcal{M}_k$ is replaced with raw utterances from session history $S_K$. This baseline allows us to measure the effectiveness of using semi-structured fine-grained preferences over raw conversational utterances.

\medskip \noindent \textbf{Recommendation Human Evaluation.}
Two human experts are instructed to evaluate the \emph{rationale} and \emph{relevance} aspects of recommendations: ``Which assistant's rationale is more in line with the user's preferences?'', and `` Which assistant's recommendation item meets user preferences?''
Following~\citet{Li:2019:AID}, we conduct pairwise comparisons between responses from our method and a baseline. For each aspect, the annotators choose win, lose, or tie options, and disagreements are resolved through discussion.
For evaluation, we randomly select 50 and 10 samples from the recipe and movie domains, respectively.

 \medskip \noindent \textbf{Recommendation Automatic Evaluation.}
Automatic machine translation measures such as ROUGE and BLEU assume significant overlap between ground truth and valid responses. This strong assumption do not hold for dialogue systems and in particular for personal recommendations, where valid responses represent high level of diversity. The lack of correlation between human evaluation is reported in previous study~\cite{Liu:2016LHNE} and has been observed in our study as well. 
Addressing this challenge, we introduce an automatic reference-based evaluation metric, \textbf{Preference Utilization (PU)}, which aims to measure the utilization of user preferences in the recommendation response.
Let $ \mathcal{P} = \{p_1, p_2, \ldots\} $ denote the set of preferences (without their corresponding categories as defined in Eq.~\ref{eq:pref}).
The subsets $ \mathcal{P}^{pred} \subseteq \mathcal{P} $ and $ \mathcal{P}^{ref} \subseteq \mathcal{P} $ denote preferences $p_i \in \mathcal{P}$ that appear in predicted and reference recommendation responses, respectively. Preference utilization precision $P_{PU}$ and recall $R_{PU}$ are computed as: 
\begin{align*}
    P_{PU} = \frac{|\mathcal{P}^{pred} \cap \mathcal{P}^{ref}|}{|\mathcal{P}^{pred}|}, 
    \quad
    R_{PU} = \frac{|\mathcal{P}^{pred} \cap \mathcal{P}^{ref}|}{|\mathcal{P}^{ref}|}.
\end{align*}
In our experiments, we perform exact string matching to generate $\mathcal{P}^{pred}$ and $\mathcal{P}^{ref}$.
When refference utterance $ u^{ref} $ includes URLs, we extract the content from the corresponding web page and use it in string matching.
For our ground truth, we only use the assistant responses about recommendations which are accepted by the user.

\medskip \noindent \textbf{Preference Extraction Evaluation.}
We evaluate preference extraction using exact match and BERTScore~\cite{Zhang:2020:BET}.
Exact match assesses the case-insensitive string match of preferences between predictions and ground truth, while BERTScore accounts for different expressions of identical preferences.
Here, preferences within each category are flattened into a single string with commas as delimiters and used for computation of BERTScore.

\subsection{Experimental Setup}

For preference extraction, we fine-tuned Flan-T5 models on our training set using the HuggingFace Transformers library~\cite{Wolf:2020:TSN}.
For both domains, the batch size is set to 8, the learning rate to 5e-5, and the AdamW optimizer is used.
Training is conducted for 10 epochs, with the best checkpoint selected based on the validation set. We ran all our experiments on a single GPU (NVIDIA A100 40GB). For a personalized recommendation, we use Llama-2-7B~\cite{Touvron:2023:LOF} due to its ability to handle a sufficiently long context, while being capable of running on a single GPU. We run the model (from the HuggingFace model hub) 10 times for each dialogue and report the average score.

\section{Results}
\label{sec:results}

This section presents the results of the proposed methods for dialogue collection (\S\ref{sec:res:dialogue}) and personal recommendation (\S\ref{subsec:Personal Recommendation Evaluation})

\begin{table}[t]
\centering
\caption{UniEval scores.
Significance against all baselines is marked by \textsuperscript{+}.
} 
\shrink
\label{tab:consolidated-uni-eval-scores}
\setlength{\tabcolsep}{6pt}
\begin{tabular}{l || c c c |c}
\hline
\textbf{Dataset} & \textbf{NAT} & \textbf{UND} & \textbf{COH} & \textbf{Avg.} \\
\Xhline{2pt}
SGD & 0.794 & 0.781 & 0.758 & 0.778 \\
M2M & 0.634 & 0.616 & 0.701 & 0.650 \\
Taskmaster-1 & 0.792 & 0.779 & 0.782 & 0.784 \\
MultiWOZ & \underline{0.870} & \underline{0.860} & 0.848 & 0.859 \\
CCPE-M & 0.716 & 0.708 & 0.689 & 0.704 \\
MG-ShopDial & 0.743 & 0.730 & 0.687 & 0.720 \\
\hline
LAPS-Recipe & 0.867 & \underline{0.860} & \underline{0.891}\textsuperscript{+} & \underline{0.872}\textsuperscript{+} \\
LAPS-Movie & \textbf{0.874} & \textbf{0.868} & \textbf{0.897}\textsuperscript{+} & \textbf{0.880}\textsuperscript{+} \\
\hline
\hline
PersonaChatGen\textsuperscript{\dag} & 0.894 & 0.887 & 0.738 & 0.839 \\
\hline
\end{tabular}
\footnotesize
\begin{flushleft}
\hspace{2.0em} \textsuperscript{\dag} Scores provided as a reference, but do not represent fair comparison.
\end{flushleft}
\end{table}

\begin{figure}[t]
\miniskip
\centering
\includegraphics[width=0.9\linewidth]{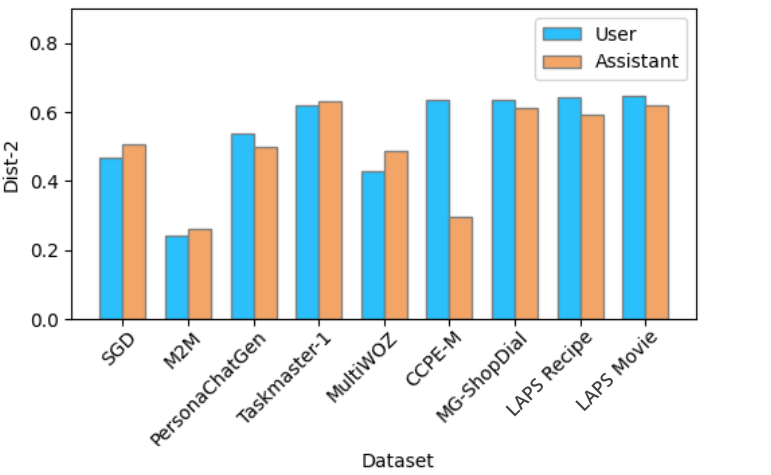}
\shrink
\caption{Lexical Diversity of user and assistant utterances.}
\label{fig:diversity-user-asst}
\shrink
\end{figure}

\subsection{Dialogue Collection Evaluation}
\label{sec:res:dialogue}

Using LAPS, we can collect a large-scale multi-session personalized dataset, as shown in Table~\ref{tab:dataset-overview}.
The number of preferences is the total number of $(s, p, o)$ triples, where $s$ is the user, $p$ is the preference category, and $o$ is the preference attribute. Using human verification, we identified  4.5\% error rate in preferences extracted by GPT-4, highlighting the need for human involvement for accurate preference extraction.


\medskip \noindent \textbf{Lexical Diversity.}
Table \ref{tab:eval-baselines} shows the results of lexical diversity evaluation,
indicating that LAPS-Movie and -Recipe achieve the highest diversity scores with respect to Dist-2 and Ent-4. Considering all metrics, LAPS-Movie is on par with MG-ShopDial, a dataset of human-human dialogues involving trained volunteers as an assistant role.
These results demonstrate that LAPS collects lexically diverse dialogues as effectively as human-human dialogue collection methods.

Comparing the lexical diversity of user and system utterances in Figure \ref{fig:diversity-user-asst}, we observe 
that our method achieves high lexical diversity for both user and assistant utterances. This suggests that LLM's guidance can help workers compose diverse responses. 
Notably, we observe that assistant utterances in CCPE-M exhibit less diversity than those of the user. 
This could be attributed to the small number of participants acting as assistants in CCPE-M and their often short and direct responses.
This highlights that achieving diversity is non-trivial, even with trained experts.
LAPS's success in achieving diversity further demonstrates the effectiveness of our method.

Table~\ref{tab:eval-synthetic} compares LAPS- and LLM-generated conversations. The results show that synthetic dialogues are less diverse than LAPS, even with GPT-4 temperature tuning. This suggests potential diversity pitfalls in synthetic personalized dialogue generation using LLMs. One way to mitigate this issue is using a synthetic persona, as in PersonaChatGen~\cite{Lee:2022:GPD}.
However, as Table \ref{tab:eval-baselines} shows, it still falls short of human-involved LAPS, suggesting the diverse nature of human preferences.

\begin{table}[t]
\centering
\shrink
\caption{Preference extraction results.
Domain adapt. represents the model is first fine-tuned on the recipe domain and then further fine-tuned on the movie domain.
}
\shrink
\label{tab:preference-extraction-recipe}
\setlength{\tabcolsep}{2.5pt}
\small
\begin{tabular}{l | c | c || ccc | ccc} 
\hline
\multirow{2}{*}{\textbf{Domain}} & \textbf{Model} & \textbf{Domain} 
& \multicolumn{3}{c |}{\textbf{Exact Match}} 
& \multicolumn{3}{c}{\textbf{BERTScore}} \\
& \textbf{Size} & \textbf{Adpt.} & P & R & F & P & R & F \\
\Xhline{2pt}
\multirow{3}{*}{Recipe} & Small & \multirow{3}{*}{---} & 0.454 & 0.408 & 0.412 & 0.590 & 0.589 & 0.589 \\
& Base & & 0.464 & 0.476 & 0.454 & 0.622 & 0.623 & 0.622 \\
& Large & & \textbf{0.532} & \textbf{0.493} & \textbf{0.494} & \textbf{0.651} & \textbf{0.650} & \textbf{0.650} \\
\hline
\hline
\multirow{2}{*}{Movie} & \multirow{2}{*}{Large} & \texttimes & 0.453 & 0.415 & 0.425 & 0.598 & 0.592 & 0.595 \\
& & \ding{51} & \textbf{0.470} & \textbf{0.424} & \textbf{0.432} & \textbf{0.618} & \textbf{0.614} & \textbf{0.616} \\
\hline
\end{tabular}
\end{table}

\begin{figure}[t]
\centering
\includegraphics[width=0.9\linewidth]{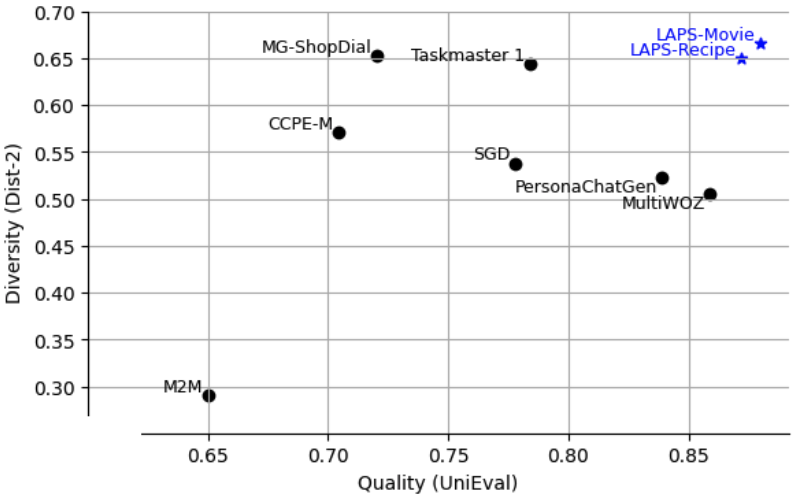}
\shrink
\caption{LAPS collects diverse and high-quality dialogues compared to other dialogue collection methods.
}
\label{fig:tmp-fig1-candidate}
\end{figure}

\medskip \noindent \textbf{Dialogue Quality.}
Table \ref{tab:consolidated-uni-eval-scores} shows the results of the dialogue quality evaluation.
For coherence, LAPS outperforms the other datasets.
For naturalness and understandability, PersonaChatGen, which is an LLM-based fully-synthetic dataset, outperforms the other datasets.
This is consistent with the findings from~\citet{Liu:2023:GNE}, which shows that LLM-based synthetic dialogues tend to have higher scores for language-model-based automatic evaluation metrics.
Excluding fully-synthetic datasets, LAPS's performance is among the best, demonstrating that our method can collect high-quality dialogues as effectively as human-human dialogue collection methods.
On average, LAPS outperforms the other datasets, highlighting the effectiveness of our method.



\medskip \noindent \textbf{Discussion.} Based on these results we can positively answer our first and second research questions: \emph{\textbf{RQ1:} Using LAPS, we can collect large-scale multi-session human-written conversations that contain actual user preferences.} and \emph{\textbf{RQ2:} LAPS-collected dialogues show high diversity and quality, on par with expert-involved human-human dialogues, as highlighted in Figure~\ref{fig:tmp-fig1-candidate}.}

\begin{table}[t]
\centering
\caption{Recommendation results.
Significance against Standard is marked by \textsuperscript{+}.
}
\shrink
\label{tab:rec-autmatic}
\begin{tabular}{l | c || c | ccc} 
\hline
\multirow{2}{*}{\textbf{Domain}} & \textbf{Prompting} & \textbf{\#Prompt} & \multicolumn{3}{c}{\textbf{\small Preference Utilization}} \\
& \textbf{Method} & \textbf{Tokens} & $\text{P}_{\text{PU}}$ & $\text{R}_{\text{PU}}$ & $\text{F}_{\text{PU}}$ \\
\Xhline{2pt}
\multirow{2}{*}{Recipe} 
& Standard & 880 & \textbf{0.554} & 0.311 & 0.398 \\
& Memory & \textbf{308} & 0.470 & \textbf{0.411}\textsuperscript{+} & \textbf{0.438}\textsuperscript{+} \\
\hline
\hline
\multirow{2}{*}{Movie} 
& Standard & 957 & \textbf{0.508} & 0.364 & \textbf{0.424} \\
& Memory & \textbf{311} & 0.443 & \textbf{0.397}\textsuperscript{+} & 0.419 \\
\hline
\end{tabular}
\shrink
\end{table}

\subsection{Personal Recommendation Evaluation}
\label{subsec:Personal Recommendation Evaluation}

\medskip \noindent \textbf{Preference Extraction.}
Table \ref{tab:preference-extraction-recipe} shows the preference extraction performance of different FlanT5 pre-trained model sizes for the recipe topic.
The results show that the performance for the Recipe domain improves as the model size increases.
Using the recipe domain as the source domain, we further fine-tune the FlanT5-Large fine-tuned model on the target movie domain.
The results show that domain adaptation consistently outperforms direct fine-tuning. This demonstrates the adaptability of our dataset to other domains through domain adaptation.



\medskip \noindent \textbf{Recommendation Human Evaluation.}
Table \ref{tab:human-eval} presents the human evaluation results for recommendation quality.
In the recipe domain, using preference memory outperforms the baseline, for both recommendation quality and rationale.
For the movie domain, the Memory method excels in rationale but not in recommendation quality. 
Given that improvements are consistently found in the rationale aspect, using preference memory is effective in improving the rationale for recommendations, a critical factor for transparent and explainable recommendations.

\medskip \noindent \textbf{Recommendation Automatic Evaluation.}
To evaluate the effectiveness of our preference-based prompting, we compare it with the baseline standard prompting method. 
The downstream recommendation task results are depicted in Table~\ref{tab:rec-autmatic}.
The Preference Utilization results show a similar trend to the human evaluation results, demonstrating the overall win of Memory over the Standard method.
In the recipe domain, Memory outperforms Standard in $\text{F}_{\text{PU}}$ score (0.438 vs. 0.398), whereas in the movie domain, their scores show no statistical significance (0.419 vs. 0.424).
Error analysis indicates that the preference memory in the movie domain is sometimes insufficient for making recommendations, suggesting a need for improving the preference extraction method.
Nevertheless, preference-based prompting for movies reduces the prompt length threefold while maintaining a comparable $\text{F}_{\text{PU}}$ score. 
This threefold reduction has significant implications for real-world applications where inference cost is a critical factor.

\begin{figure}[t]
\centering
\includegraphics[width=0.65\linewidth]{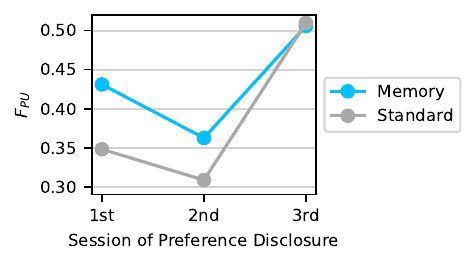}
\shrink
\caption{
    Breakdown of Preference Utilization ($\text{F}_{\text{PU}}$, recipe domain).
}
\label{fig:preference-utilization-breakdown}
\shrink
\end{figure}

\medskip \noindent \textbf{Preference Utilization by Session.}
Figure \ref{fig:preference-utilization-breakdown} shows the Preference Utilization scores by session in recipe recommendations. 
The x-axis represents the session where preferences are disclosed, as detailed in Section \ref{sec:downstream:eval-metric}.
The analysis focuses on the recommendations in the third session to examine Preference Utilization across different sessions.
The graph shows methods' struggle with utilizing preferences from earlier sessions (1st and 2nd) than those from the ongoing (3rd) session.
A similar phenomenon is reported in~\cite{Liu:2023:LIM} in a retrieval augmentation setting, referred to as \emph{LLMs' recall issue in long prompt inputs}.
The graph also demonstrates that preference-based prompting more effectively utilizes earlier session preferences than standard prompting, suggesting that preference memory can mitigate long prompt recall issues. 
We observe that this pattern (recall issues in earlier sessions and the effectiveness of preference-based prompting in addressing them) is generally consistent across the movie domain and different sessions, though not always statistically significant. Overall, our analysis shows the effectiveness of preference memory in long prompts.

\medskip \noindent \textbf{Discussion.}
Based on these results we can positively answer our last research questions: \emph{\textbf{RQ3}
Preference memory enhances effective utilization of user preferences in recommendations, improves the rationale of recommendations, and mitigates long prompt recall issues.}


\begin{table}[t]
\centering
\caption{Human evaluation results of recommendations.}
\shrink
\label{tab:human-eval}
\setlength{\tabcolsep}{4pt}
\begin{tabular}{l | c || ccc | ccc} 
\hline
\multirow{2}{*}{\textbf{Domain}} & \textbf{Prompting} & \multicolumn{3}{c|}{\textbf{\small Rationale}} & \multicolumn{3}{c}{\textbf{\small Relevance}} \\
& \textbf{Method} & Win & Lose & Tie & Win & Lose & Tie \\
\Xhline{2pt}
\multirow{2}{*}{Recipe} 
& Standard & 17 & 29 & 4 & 18 & 22 & 10 \\
& Memory & \textbf{29} & \textbf{17} & 4 & \textbf{22} & \textbf{18} & 10 \\
\hline
\hline
\multirow{2}{*}{Movie} 
& Standard & 3 & 6 & 1 & \textbf{4} & \textbf{3} & 3 \\
& Memory & \textbf{6} & \textbf{3} & 1 & 3 & 4 & 3 \\
\hline
\end{tabular}
\shrink
\end{table}

\section{Conclusion}

In this research, we proposed a method to collect large-scale multi-session personalized conversations reflecting actual user preferences.
Our method, LAPS, employs LLMs to generate personalized guidance for human workers, reducing the cognitive load for a highly complex task.
Extensive experiments demonstrate, while being a scalable and high-quality data collection method, LAPS can collect utterances as diverse as the expert-involved methods.
We further showed that utilizing extracted user preferences results in more effective personal recommendations compared to using raw user utterances of previous sessions.
In our experiment, fully-synthetic LLM-based methods does not yield diverse conversations. Using actual user preferences from LAPS as personas  is a promising avenue to explore for future.



\medskip \noindent \textbf{Acknowledgments.}
This work is supported by the Rad\-boud-Glas\-gow Collaboration Fund and in part by the Engineering and Physical Sciences Research Council (EPSRC) Grant EP/V025708/1.

%
%

\balance
\bibliographystyle{ACM-Reference-Format}
\bibliography{references-base}

\end{document}